\documentclass{cjpsuppl}
\usepackage{epsfig}

\def\sl#1{\slash{\hspace{-0.2 truecm}#1}}
\renewcommand{\d}{{\rm d}}

\def\be{\begin{equation}}
\def\ee{\end{equation}}
\def\bea{\begin{eqnarray}}
\def\eea{\end{eqnarray}}
\def\oneh{{\textstyle {1\over 2}}}

\begin{document}
 \slacs{.6mm}
 \title{Chiral-odd generalized parton distributions and transversity in light-front constituent quark models}
 \authori{M.~Pincetti, B.~Pasquini, S.~Boffi}
 \addressi{Dipartimento di Fisica Nucleare e Teorica, Universit\`a degli
Studi di Pavia and\\  INFN, Sezione di Pavia, Pavia, Italy}
 \authorii{}    \addressii{}
 \authoriii{}   \addressiii{}
 \authoriv{}    \addressiv{}
 \authorv{}     \addressv{}
 \authorvi{}    \addressvi{}
 \headtitle{ \ldots}
 \headauthor{M.~Pincetti, B.~Pasquini, S.~Boffi}
 \lastevenhead{author: title \ldots}
 \refnum{}
 \daterec{} 
 \suppl{?}  \year{2006} \setcounter{page}{1}
 \maketitle

 \begin{abstract}
 We present the general framework to calculate chiral-odd generalized parton distributions in the overlap representation using the Fock-state decomposition in the transverse-spin basis. In the forward limit we derive the transversity distribution, the tensor charge and the angular momentum sum rule for quarks with transverse polarization in an unpolarized nucleon. Numerical results are obtained by applying the formalism to the case of light-cone wavefunctions of constituent quark models.
 \end{abstract}
 

 \section{Introduction} 
In high-energy processes the quark-gluon structure of the nucleon is described by a set of parton distributions. At the level of leading twist a complete quark-parton model of the nucleon requires three quark distributions, i.e. the quark density, or unpolarized distribution, $f_1(x)$, the longitudinal polarization, or helicity, distribution, $g_1(x)$, and the transverse polarization, or transversity, distribution, $h_1(x)$. The first two distributions are well known quantities and can be extracted from inclusive deep-inelastic scattering (DIS) data. The last one is totally unknown because, being a chiral-odd quantity, does not contribute to inclusive DIS and is only accessible experimentally when coupled to another chiral-odd partner in the cross section. 

Theoretically, these quark distributions are the forward limit of so-called generalized parton distributions (GPDs) defined as non-diagonal hadronic matrix elements of bilocal products of the light-front quark and gluon field operators. They depend on the momentum transferred to the parton, as well as on the average longitudinal momentum. A complete set of quark GPDs at leading twist include four helicity conserving, usually labeled $H$, $E$, $\tilde H$, $\tilde E$, and four helicity flipping (chiral-odd) GPDs, labeled $H_T$, $E_T$, $\tilde H_T$, $\tilde E_T$~\cite{HJ98,diehl01}. In the forward limit, as the momentum transfer vanish, $H$, $\tilde H$ and $H_T$ reduce to $f_1$, $g_1$ and $h_1$, respectively. 

A variety of model calculations is available for the helicity conserving GPDs~\cite{goeke,diehl03,ji04,BR05}. Less attention has been paid up to now to the chiral-odd case. Up to now, there is only one proposal to access 
the chiral-odd GPDs in diffractive double meson production~\cite{Pire1,Pire2}. 
In the present paper the chiral-odd GPDs are studied in the overlap representation of light-cone wave functions (LCWFs) that was originally proposed in Refs.~\cite{diehl2,brodsky} within the framework of light-cone quantization. In a fully covariant approach the connection between the overlap representation of GPDs and the non-diagonal one-body density matrix in momentum space has further been explored in Ref.~\cite{BPT03} making use of the correct transformation of the wave functions from the  (canonical) 
instant-form to the (light-cone) front-form description. In this way the lowest-order Fock-space components of LCWFs with three valence quarks are directly linked to wave functions derived in constituent quark models (CQMs). Results for the four helicity conserving GPDs have been 
obtained~\cite{BPT03,BPT04}, automatically fulfilling the support condition and the particle number and momentum sum rules. This approach has been extended to include the next-order Fock state component, by developing a convolution formalism for the unpolarized GPDs which incorporatae the quark sea contribution~\cite{BTB05,PB06}. Important dynamical effects are introduced by the correct relativistic treatment; as a consequence, e.g., a nonzero anomalous magnetic moment of the nucleon is obtained even when all the valence quarks are accommodated in the $s$-wave. An effective angular momentum, as required by the arguments of Refs.~\cite{BHMS,burkardt05}, is introduced by the boost from the rest frame to the light-front frame producing a nonvanishing unpolarized nonsinglet (helicity-flip) quark distribution.

In this paper, following the approach discussed in Ref.~\cite{BBP05}, chiral-odd GPDs are derived using the Fock-state decomposition in the transverse-spin basis and LCWFs with three valence quarks. Numerical results are presented including the forward limit, i.e. transversity, and its first moment, i.e. the tensor charge.


 \section{The overlap representation for the chiral-odd GPDs}

The chiral-odd GPDs are off-diagonal in the parton helicity basis. Working in the reference frame where  the momenta $\vec p$ and $\vec p\,'$ of the initial and final nucleon lie in the $x-z$ plane, they become diagonal if one changes basis from eigenstates of helicity to eigenstates of transversity, i.e. states with spin projection $\uparrow$ ($\downarrow$) directed along (opposite to) the transverse direction $\hat x$. In such a basis they are obtained by means of the following relations
\begin{eqnarray}
H^{q}_T &=& \frac{1}{\sqrt{1 - \xi^2}}T^q_{\uparrow\uparrow} -
\frac{2M\xi} {\epsilon\sqrt{t_0 - t}(1 - \xi^2)}T^q_{\uparrow\downarrow},
\label{eq:accat}\\
E_T^{q} &=& 
\frac{2M\xi}{\epsilon\sqrt{t_0 - t}}\frac{1}{1 - \xi^2}T_{\uparrow\downarrow}^q
+ \frac{2M}{\epsilon\sqrt{t_0 - t}(1 - \xi^2)}\tilde{T}^q_{\uparrow\uparrow}
\nonumber\\
&&- \frac{4M^2}{(t_0 - t)\sqrt{1 - \xi^2}(1 - \xi^2)}
\bigg(\tilde{T}^q_{\downarrow\uparrow} - T^q_{\uparrow\uparrow}\bigg),
\label{eq:et}\\
\tilde{H}_T^{q} &=& \frac{2M^2}{(t_0 - t)\sqrt{1 -\xi^2}}
(\tilde{T}^q_{\downarrow\uparrow} - T^q_{\uparrow\uparrow}),
\label{eq:tildeaccat}\\
\tilde{E}^{q}_T &=& \frac{2M}{\epsilon\sqrt{t_0 - t}(1 -\xi^2)}
\bigg(T^q_{\uparrow\downarrow}
+ \xi\tilde{T}^q_{\uparrow\uparrow}\bigg)\nonumber\\ &&
- \frac{4M^2\xi}{(t_0 - t)\sqrt{1 - \xi^2}(1 - \xi^2)}
\bigg(\tilde{T}^q_{\downarrow\uparrow} - T^q_{\uparrow\uparrow}\bigg),
\label{eq:tldeet}
\end{eqnarray}
where $t=\Delta^2$ is the invariant momentum square, $- t_0 = 4 m^2 \xi^2/(1-\xi^2)$ is its minimum value for a given value of the skewness parameter $\xi=-\Delta^+/2P^+$ defined in terms of the momentum transfer $\Delta^\mu=p'^\mu-p^\mu$ and the average nucleon momentum $P^\mu=\oneh(p+p')^\mu$, and $\epsilon = \mathrm{sign}(D^1)$, where $D^1$ is the $x$-component of $D^\alpha = P^+ \Delta^\alpha - \Delta^+ P^\alpha$.

The transverse matrix elements are defined as follows
\begin{eqnarray}
\label{eq:T}
T^{q}_{\lambda'_t\lambda_t} &=& \langle p',
\lambda'_t|\int\frac{dz^-}{2\pi}e^{i \bar xP^+z^-}
\bar{\psi}(-{z}/2)\gamma^+\gamma^1\gamma_5
\psi({z}/2)|p, \lambda_t\rangle,
\\
\label{eq:Ttilde}
\tilde{T}^{q}_{\lambda'_t\lambda_t} &=&\langle p',
\lambda'_t|\int\frac{dz^-}{2\pi}e^{i\bar xP^+z^-}
\frac{i}{2}\bar{\psi}(-{z}/2)\sigma^{+1}\psi({z}/2)
|p, \lambda_t\rangle.
\end{eqnarray}
In the region $\xi \leq \bar x\leq 1$ of plus-momentum fractions, where the generalized quark distributions describe the emission of a quark with plus-momentum $(\bar x+\xi)P^+$ and its reabsorption with plus-momentum $(\bar x-\xi)P^+$, they can be written in the overlap representation of LCWFs as
\begin{eqnarray}
\label{eq:T_ov}
T^{q }_{\lambda'_t\lambda_t} &=&
\sum_{N,\beta=\beta'}
\bigg(\sqrt{1 - \xi}\bigg)^{2 - N}\bigg(\sqrt{1 + \xi}\bigg)^{2 -
N}
\sum_{j=1}^N \mathrm{sign}(\mu_{j}^{t})
\delta_{s_jq}\nonumber\\
&&\times\int[d\bar{x}]_N[d^2\vec{k}_\perp]_N\delta(\bar{x} -
\bar{x}_j)\mathit{\Psi}^{*}_{\lambda'_t,N,\beta'}(\hat{r}')
\mathit{\Psi}_{\lambda_t,N,\beta}(\tilde{r}),
\label{eq:tuno}\\
& &\nonumber\\
\label{eq:Ttilde_ov}
\tilde{T}^{q }_{\lambda'_t\lambda_t} &=&
\sum_{\beta,\beta',N}
\bigg(\sqrt{1 - \xi}\bigg)^{2 - N}\bigg(\sqrt{1 + \xi}\bigg)^{2 -
N}
\sum_{j=1}^N\delta_{{\mu_j^t}'-\mu_j^t}\delta_{{\mu_i^t}'\mu_i^t}\mathrm{sign}(\mu_{j}^{t})
\delta_{s_jq}\nonumber\\
&&\times\int[d\bar{x}]_N[d^2\vec{k}_\perp]_N\delta(\bar{x} -
\bar{x}_j)\mathit{\Psi}^{*}_{\lambda'_t,N,\beta'}(\hat{r}')
\mathit{\Psi}_{\lambda_t,N,\beta}(\tilde{r}),
\label{eq:tdue}
\end{eqnarray}
where $\Psi_{\lambda_t ,N,\beta}$ is the momentum LCWF of the $N$-parton Fock state, $s_j$ labels the quantum numbers of the $j$-th active parton, with transverse initial (final) spin polarization $\mu^{t}_{j}$ (${\mu_j^t}'$),  and $\mu^{t}_{i} \, ({\mu_i^t}')$ are the transverse spin of the spectator initial (final) quarks. The set of kinematical variables $r,r'$ are defined according to Refs.~\cite{diehl2,BBP05}. The integration measures in Eqs.~(\ref{eq:tuno}) and (\ref{eq:tdue}) are defined as
\be
[\d x]_N \! = \! \prod_{i=1}^N dx_i \,\delta\!\left(1-\sum_{i=1}^N x_i\right)\!, \ 
 [\d^2 {\vec k}_\perp]_N \! = \! \frac{1}{(16\pi^3)^{N-1}}\prod_{i=1}^N\delta^2\!\left(\sum_{i=1}^N{\vec k}_{\perp,i} - {\vec p}_\perp\right)\! d^2\vec k_{\perp i}.
 \ee


 \section{Results}

As an application of the general formalism reviewed in the previous sections we consider the valence-quark contribution ($N=3$) to the chiral-odd GPDs calculated starting from an instant-form SU(6) symmetric wave function of the proton derived in the relativistic quark model of Ref.~\cite{Schlumpf94a}. The structure of the nucleon wave function in this model is SU(6) symmetric for the spin-isospin components. In such a model, e.g., from the forward limit of the chiral-even GPDs the anomalous magnetic moments for the $u$ and $d$ quarks are $\kappa^u=1.88$ and $\kappa^d=-1.59$, respectively, corresponding to fairly good proton and neutron anomalous magnetic momenta $\kappa^p=1.78$ and $\kappa^n=-1.68$.

\begin{figure}[ht]
\begin{center}
\epsfig{file=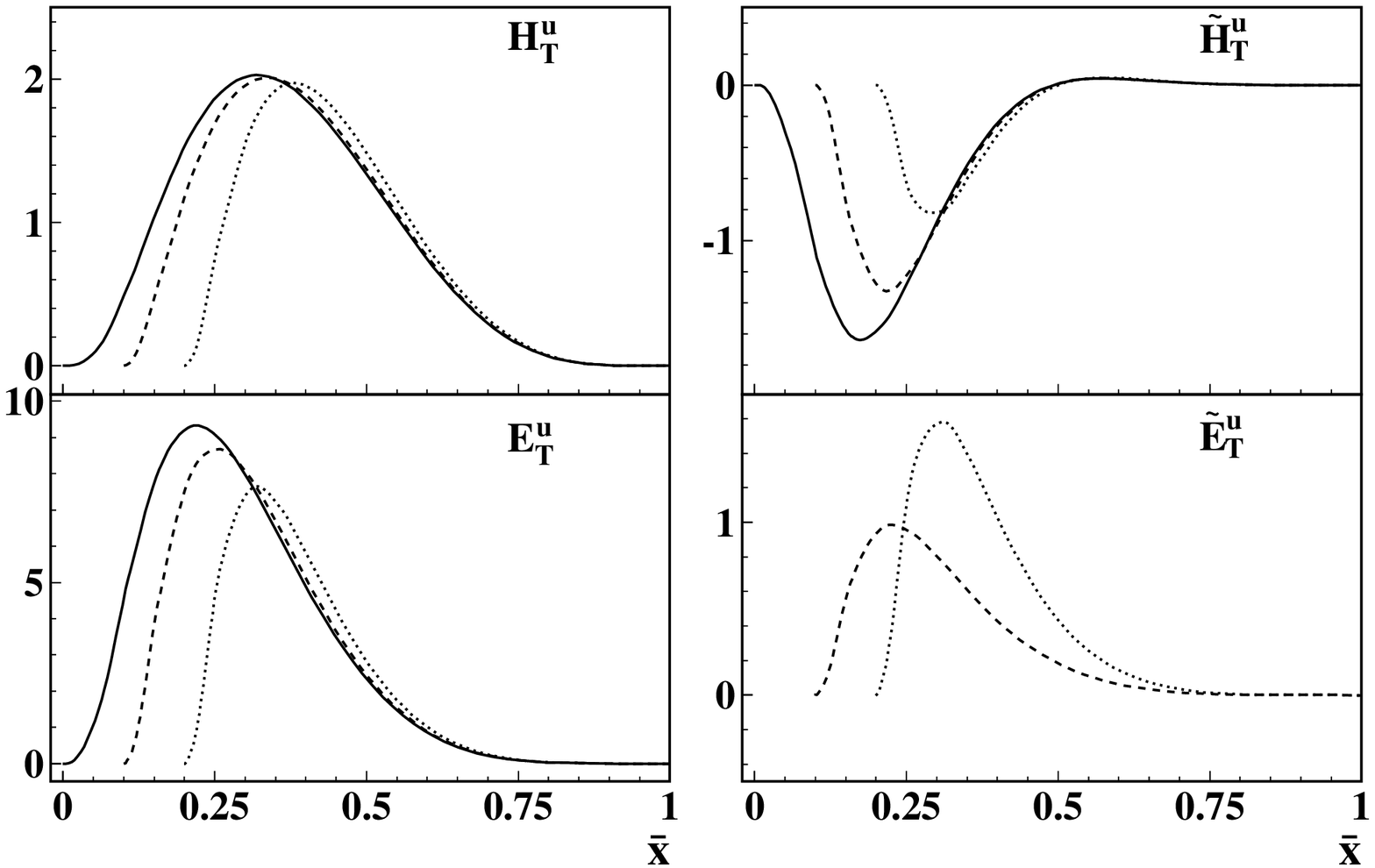,  width=24 pc}
\epsfig{file=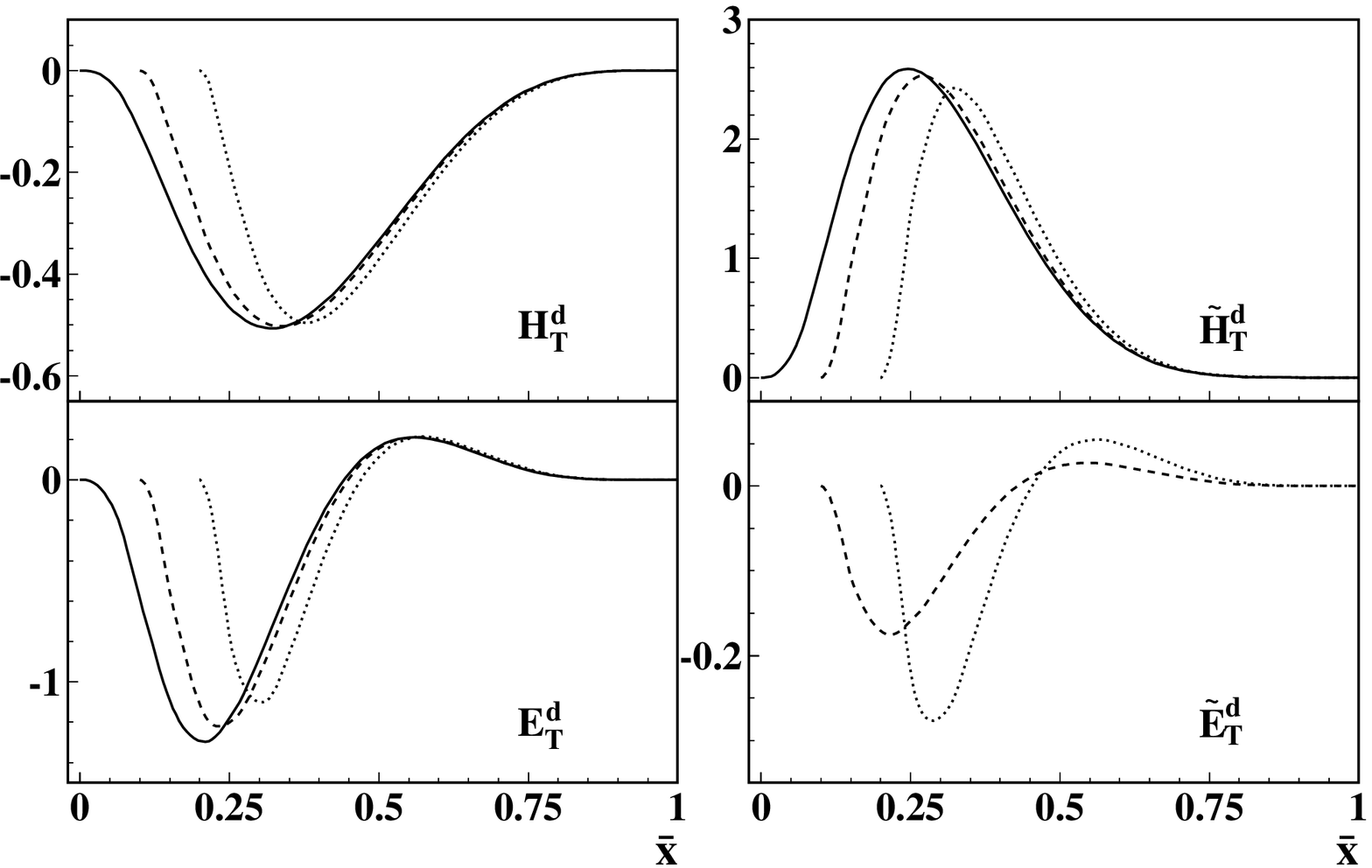,  width=24 pc}
\end{center}
\vspace{-0.4cm}
\caption{\small The chiral odd generalized parton distributions calculated in the hypercentral CQM for the flavours $u$ and $d$ at $t=-0.2$ (GeV)$^2$ and for different values of $\xi$: $\xi =0$ (solid curves), $\xi=0.1$ (dashed curves), $\xi=0.2$ (dotted curves).}
\label{fig:fig1}
\end{figure}

As an example we show in Fig.~\ref{fig:fig1} the four calculated chiral-odd GPDs, $H^q_T$, $E^q_T$, $\tilde H^q_T$, $\tilde E^q_T$ at $t=-0.2$ (GeV)$^2$ and for different values of $\xi$. In all cases the GPDs vanish at $\bar x=\xi$ since in our approach they include the contribution of valence quarks only and we cannot populate the so-called ERBL region with $\vert\bar x\vert\le \xi$ where quark-antiquark pairs and gluons are important. Therefore, at low $\bar x$ this gives a strong $\xi$ dependence of the peak position of the distribution, but for large $\bar x$ the $\xi$ dependence turns out to be rather weak.

Concerning the $t$ dependence, it affects the low-$\bar x$ region and is more pronounced in the cases of $E^q_T$ and $\tilde H^q_T$. For large $\bar x$ values the decay of all the distributions towards zero at the boundary $\bar x=1$ is almost independent of $t$ (see Ref.~\cite{BBP05}).

\section{The forward limit and the tensor charge}

In the forward limit $\Delta^\mu\rightarrow 0$ ($\bar x\to x$, with $x$ being the usual Bjorken variable), only the quark GPDs $H_T^q$ can be measured and, in fact, becomes the quark transversity distributions $h_1^q(x)$. Although the quark GPDs $E_T^q$ and $\tilde H_T^q$ do not contribute to the scattering amplitude, they remain finite in the forward limit, whereas $\tilde E_T^q$ vanishes identically being an odd function of $\xi$ as already noticed~\cite{diehl01}.

The transversity distribution $h^q_1$ is the counterpart in the transverse-polarization space
of the helicity parton distribution $g^q_1$ which measures the helicity asymmetry. As it was stressed by Jaffe and Ji~\cite{jaffe91}, in nonrelativistic situations where rotational and boost operations commute, 
one has $g^q_1=h^q_1$. Therefore the difference between $h^q_1$ and $g^q_1$  is a measure of the relativistic nature of the quarks inside the nucleon. In light-cone CQMs these relativistic effects are encoded in the Melosh rotations.

\vspace{-1cm}
\begin{figure}[h]
\begin{center}
\epsfig{file=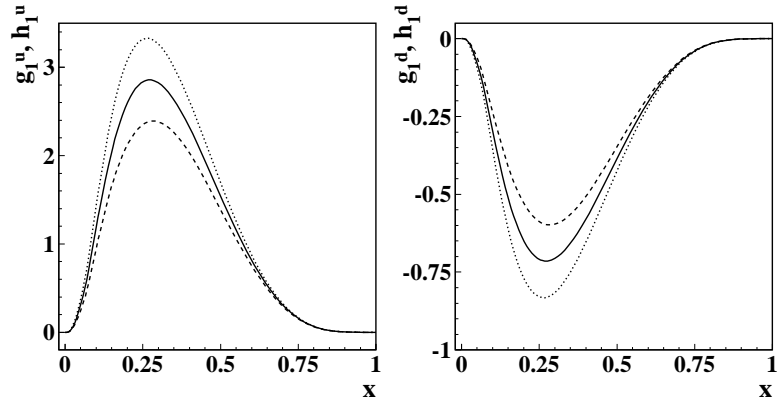,  width=25 pc}
\end{center}
\caption{\small Helicity and transversity distributions for the $u$ (left panel) and $d$ (right panel) quark. The solid lines correspond to $h^q_1$, the dashed lines show $g^q_1$, and the dotted lines are the nonrelativistic results when Melosh rotations reduce to the identity ($h^q_1=g^q_1$).}
\label{fig:fig2}
\end{figure}

In Fig.~\ref{fig:fig2} the helicity and transversity distributions, $g_1$ and $h_1$, obtained as a forward limit of the corresponding GPDs calculated with the CQM of Ref.~\cite{Schlumpf94a} are compared and plotted together with the nonrelativistic result when Melosh rotations reduce to identity. The large difference between $g_1$ and $h_1$ shows how big is the effect of relativity. 

Recalling the expression for the unpolarized parton distribution $f^q_1$ obtained in Ref.~\cite{BPT03}, it is easy to see that the following relations hold
\begin{eqnarray}
\label{eq:soffer}
2h^u_1(x)=g^u_1(x)+\frac{2}{3}f^u_1(x),
& \quad&
2h^d_1(x)=g^d_1(x)-\frac{1}{3}f^d_1(x),
\end{eqnarray}
which are compatible with the Soffer inequality~\cite{soffer}. In the nonrelativistic limit one obtains  $h^u_1=g^u_1=2/3 f^u_1$ and $h^d_1=g^d_1=-1/3 f^d_1.$ We note that the relations (\ref{eq:soffer}) generalize to the case of parton distributions the results obtained in Refs.~\cite{SS97,MaSS98} for the axial ($\Delta q$) and tensor ($\delta q$) charges, defined as
\begin{equation}
\Delta q=\int_{-1}^1 d x \, g^q_1(x),\label{eq:axialcharge} \quad
\delta q=\int_{-1}^1 d x \, h^q_1(x),\label{eq:tensorcharge}
\end{equation}
respectively. The tensor (axial) charge measures the net number of transversely (longitudinally) polarized valence quarks in a transversely (longitudinally) polarized nucleon. As shown in Table~\ref{table:tab1} the up-quark contribution dominates in a transversely polarized proton. However, the obtained numbers are closer to those derived in the nonrelativistic approach than those predicted by QCD sum rules~\cite{HeJi96} ($\delta u = 1.33\pm 0.53$ and $\delta d = 0.04\pm 0.02$ at the scale of the nucleon mass), with a nonnegligible negative contribution of the down quark. 
When evolved in leading-order QCD from the intrinsic scale of the model ($Q_0^2=0.079$ GeV$^2$) to $Q^2=10$ GeV$^2$ the tensor charges become $\delta u =0.79$ and $\delta d=-0.20$ in the CQM of Ref.~\cite{Schlumpf94a}, within the range of values calculated in the different models considered in Ref.~\cite{barone} and in fair agreement with lattice QCD calculations~\cite{alikhan}.

\begin{table}[h]
\begin{center}
\caption{Valence contributions to the axial and tensor charge calculated within different SU(6)-symmetric quark models: the nonrelativistic quark model (NR), the harmonic oscillator model (HO) of Ref.~\cite{SS97}, and the CQM model of Ref.~\cite{Schlumpf94a}.}
\vspace{2mm}
\begin{tabular}{|crrr|}
\hline
{} & $\, $ NR$\, $ &$\, $ HO$\, $ & $\, $\cite{Schlumpf94a} $\, $  \\ 
\hline 
$\Delta u\,$ & $\,4/3\,$  & $\,1.00\,$ & $\,1.00\,$ \\
$\Delta d\,$ & $\,-1/3\,$  & $\,-0.25\,$ & $\,-0.25\,$ \\
$\delta u\,$ & $\,4/3\,$     & $\,1.17\,$  & $\,1.16\,$ \\
$\delta d\,$ & $\,-1/3\,$ & $\,-0.29\,$ &$\,-0.29\,$ \\
\hline
\end{tabular}
\end{center}
\label{table:tab1}
\end{table}

Another quantity related to the forward limit of chiral-odd GPDs is the angular momentum $J^x$ carried by quarks with transverse polarization in the $\hat x$ direction in an unpolarized nucleon at rest. This quantity has recently been shown~\cite{Bur05} to be one half of the expectation value of the 
transversity asymmetry, i.e.
\be
\langle \delta^x J^x_q\rangle = \langle J^x_{q,+\hat x} - J^x_{q,-\hat x}\rangle =
\oneh\left[ A_{T20}(0) +2\tilde A_{T20} (0) + B_{T20}(0) \right],
\ee
where the invariant form factors $ A_{T20}(t),$ $\tilde A_{T20}(t)$ and $B_{T20}(t)$ are the second moments of the chiral-odd GPDs $H_T$, $\tilde H_T$ and $E_T$, respectively~\cite{diehl05,Bur05}. Using LCWFs derived from the relativistic CQM of Ref.~\cite{Schlumpf94a} we obtain 
\be
 \langle \delta ^x J^x_u\rangle=0.54, 
\quad \langle \delta ^x J^x_d\rangle=0.37,
\qquad\qquad \mbox{\cite{Schlumpf94a}}
\ee
while using the simple harmonic oscillator wave function of the nucleon as in Ref.~\cite{SS97}, we obtain 
\be
 \langle \delta ^x J^x_u\rangle=0.68, \quad \langle \delta ^x J^x_d\rangle=0.28.
\qquad\qquad \mbox{(HO)}
\ee

The same also occurs for the forward matrix element of $2\tilde H_T + E_T$, i.e.
\be
\kappa^q_T \equiv \int dx\left[2\tilde H^q_T(x,0,0) + E^q_T(x,0,0)\right].
\label{eq:kappat}
\ee
The quantity $\kappa^q_T$  describes how far and in which direction the average position of quarks with spin in the $\hat x$ direction is shifted in the $\hat y$ direction in an unpolarized nucleon~\cite{Bur05}. Thus $\kappa^q_T$ governs the transverse spin-flavor dipole moment in an unpolarized nucleon and plays a role similar to the anomalous magnetic moment $\kappa^q$ for the unpolarized quark distributions in a transversely polarized nucleon. As a matter of fact, we obtain
\be
\begin{array}{cclccl}
\kappa^u_T & = &  3.98, \quad \kappa^d_T & = & 2.60, \qquad\qquad & \mbox{(\cite{Schlumpf94a})}\\
\kappa^u_T & = &  3.60, \quad \kappa^d_T & = & 2.36. \qquad\qquad & \mbox{(HO)}
\end{array}
\ee
Apart from their magnitude, the same sign of $\kappa^q_T$ is predicted in both models. This may have an impact on the Boer-Mulders function  $h_1^{\perp q}$ describing the asymmetry of the transverse momentum of quarks perpendicular to the quark spin in an unpolarized nucleon~\cite{BM98}. Since for $\kappa_T>0$ we expect that quarks polarized in the $\hat y$ direction should preferentially be deflected in the $\hat x$ direction, in accordance with the Trento convention~\cite{TN04} $\kappa^q_T>0$ would imply $h_1^{\perp q}<0$~\cite{Bur05}. Furthermore, keeping in mind the magnitude of the quark anomalous magnetic moments $\kappa^q$ derived with relativistic CQM of Ref.~\cite{Schlumpf94a}, the average Boer-Mulders function is predicted here larger than the average Sivers function~\cite{Sivers}, $f_{1T}^{\perp q}\sim -\kappa^q$, describing the transverse momentum asymmetry of unpolarized quarks in a transversely polarized target.

Work is in progress to calculate the transverse double-spin asymmetry in Drell-Yan lepton-pair production with properly $Q^2$-evoluted transversity.

 \bigskip

 {\small 
This research is part of the EU Integrated Infrastructure Initiative Hadronphysics Project under contract number RII3-CT-2004-506078.
}

 \bigskip


 \end{document}